\documentstyle[12pt,world_sci]{article}
\begin{document}

\title{{\bf CHARM PRODUCTION AT RHIC AND LHC TO $O(\alpha_{s}^3)$}}
\author{INA SARCEVIC\\
{\em Department of Physics, University of Arizona, Tucson,\\
AZ 85721, USA}}\\
\maketitle
\setlength{\baselineskip}{2.6ex}
\vspace{1.0cm}
\begin{center}
\parbox{13.0cm}
{\begin{center} ABSTRACT \end{center}
{\small \hspace*{0.3cm}
We present results on rapidity and transverse momentum
distributions of inclusive charm quark production in
hadronic and heavy-ion collisions at RHIC and LHC energies,
including the next-to-leading
order, $O(\alpha_s^3)$, radiative corrections and the
nuclear shadowing effect.$^1$
We determine hadronic and nuclear
K-factor
for the differential and total inclusive cross sections for
the charm production.
We discuss theoretical uncertainties inherent in our calculation.
In particular, we find that
different extrapolations of
gluon density
to low-$x$ region
introduce large theoretical uncertainty in the calculation of
charm production at LHC energies.
Finally, we comment on the possibility
of detecting the quark-gluon plasma signal as an
enhanced charm production in heavy-ion collisions at RHIC.

}}
\end{center}

\section{Introduction}

The main goal of the future heavy-ion colliders, such as
RHIC and LHC, is to
study the properties of nuclear matter under extreme conditions, and
in particular to search for  the formation of  a
new state of matter, the
quark-gluon plasma.$^2$
With the assumption
that high energy nuclear collisions lead to the
thermalized system,
RHIC and LHC energies are sufficiently
large to
produce very dense matter,
of the order of several $GeV/fm^{3}$,
well above the critical density necessary to create
the quark-gluon plasma
.$^3$  The problem of finding a clean, detectable signal
for this phase is currently one of the most challenging theoretical
problems.
In the last few years, the possible signals, such as
thermal photons, dileptons and $J/\Psi$ suppression,
have been
investigated in detail and found to have yields comparable in magnitude
with those expected from a simple
extrapolation of hadronic collisions.$^4$
Recently,  open charm production in heavy-ion collisions
has been proposed as an
elegant method for probing the possible formation of the
quark-gluon plasma.$^5$
However, in order
to determine whether the enchanced charm production at RHIC can be
unambiguously interpreted as a signal of QGP,
one needs to have control of the other sources of
charm production.  The standard competing
process is  charm production through the hard collisions
of partons inside the nuclei.
In addition to being relevant as
a background for the signal of quark-gluon plasma, this
type of charm production is useful tool for studying
the perturbative aspects of strong interactions and for
determining
the nuclear screening/shadowing effect on the gluon
distribution in a nucleus.

\section{Charm Production at RHIC and LHC Energies}

Here we present results of  calculations of the
rapidity and transverse momentum distributions of
inclusive charm quark production in
p-p and Au-Au collisions at RHIC and LHC,
including the $O(\alpha_s^3)$
radiative
corrections and the
nuclear shadowing effect.
We start by giving the expression for the
differential inclusive distribution of
charm production
in
nuclear collisions, which in perturbative QCD
is obtained by
convolution of parton densities in nuclei with a hard scattering
parton cross
section.
For the parton cross sections,
we include leading-order subprocesses, $O(\alpha_s^2)$, such as
$ q+\bar q\rightarrow Q+\bar Q$ and
$ g+g\rightarrow Q+\bar Q$,
and next-to-leading order contributions, $O(\alpha_s^3)$, such as
$ q+\bar q\rightarrow Q+\bar Q +g$,
$ g+q \rightarrow Q+\bar Q+g$,
$ g+\bar q \rightarrow Q+\bar Q+\bar q$ and
$ g+g \rightarrow Q+\bar Q+g $.  The double differential inclusive
distribution of
charm production in central
AA collisions can be written as
\begin{equation}
\frac{d N_c}{d^{2}p_{T}dy}=T_{AA}(0)
\frac{d{\sigma}_{c}
}{d^{2}p_{T}dy},
\end{equation}
where the double differential inclusive cross section
is given by
\begin{equation}
\frac{d{\sigma}_{c}
}{d^{2}p_{T}dy}
=
\sum_{i,j}^{partons}
\int dx_{a} dx_{b}
F_{i}^{A}(x_{a},Q^{2})F_{j}^{A}(x_{b},Q^{2})
\frac{d\hat{\sigma}_{i,j}
(Q^2, m_c, \hat s)}{d^{2}p_{T}dy},
\end{equation}
and $T_{AA}(0)$ is the nuclear overlapping density at zero impact
parameter,
$F_{i}^{A}(x,Q^2)$ is the parton structure function in a
nucleus,
$x_a$ and
$x_b$ are the fractional momenta of the incoming partons,
$\hat s$ is the parton-parton c.m. energy ($\hat s=x_a x_b s$).
The parton differential cross section calculated to
$O(\alpha_{s}^{3})$ and the next-to-leading order expression
for the coupling constant $\alpha_s(Q^2)$ can be found in
Ref. 6.

To obtain the number of nucleon-nucleon collisions
per unit of transverse
area at fixed impact parameter, we consider the
nuclear overlapping function$^7$
\begin{equation}
T_{AA}(b)=\int d^{2}b_{1}
T_{A}(\mid\vec{b}_{1}\mid) T_{A}(\mid\vec{b} - \vec{b}_{1}\mid),
\end{equation}
where
$T_A(b)$ is
the nuclear profile function (i.e.
the nuclear density integrated over the longitudinal size).
For the
nuclear density we take the
Woods-Saxon
distribution.$^8$
For central collisions
the
overlapping function  can be approximated
by $T_{AA}(0)=A^2/\pi R_A^2$, which gives
$T_{Au-Au}(0)=30.7 mb^{-1}$.

\section{The Nuclear Shadowing Effect}

\par
The nuclear parton distribution, if
nucleons were independent, would be given
as A times the parton structure function in a nucleon.
However, at high
energies, the parton densities become so large that the sea quarks and
gluons overlap
spatially and the nucleus can not be viewed as a collection of uncorrelated
nucleons.
This happens when the longitudinal size of the parton, in
the infinite momentum frame of the nucleus, becomes larger than the
size of the nucleon.  Partons from different nucleons start
to interact and through annihilation effectively reduce the parton
density in a nucleus.  When partons inside the nucleus completely
overlap, there reach a saturation point.  Motivated by this simple
parton picture of the nuclear shadowing effect and taking into account
the $A^{1/3}$
dependence obtained by considering the modified, nonlinear
Altarelli-Parisi equations with gluon recombination included, the
modifying factor to the
parton structure function in a nucleus can be written as$^9$
\begin{equation}
R(x,A)\equiv \frac {F_{i}^A(x,Q^2)}{A F_{i}^N(x,Q^2)}
=\left\{\begin{array}{ll}
1-\frac{3}{16}x+\frac{3}{80}&.2<x\leq 1\\
1&x_{n}<x\leq .2\\
1-D(A^{1/3}-1)\frac{1/x-1/x_{n}}{1/x_{A}-1/x_{n}}&x_{A}\leq x\leq x_{n}\\
1-D(A^{1/3}-1)&0<x<x_{A}
\end{array}
\right
\end{equation}
where
$F_{i}^N(x,Q^2)$ is the parton structure function in a nucleon,
$x_{n}=1/(2r_{p}m_{p})$,
$x_A$ is a saturation point ($x_{A}=1/(2R_{A}m_{p})$),
$m_p$ is the mass of the proton,
$r_{p}$ is the radius of a proton and
$R_{A}$ is the
radius of the nucleus.
By fitting the
parameter $D$ to
all
deep inelastic lepton-nucleus data on the
ratio $F_2^A(x,Q^2)/F_2^D(x,Q^2)$$^{10}$
we
find that $R(x,A)$ has
much steeper x-dependence than the data, especially
for
$0.002\leq x \leq 0.1$, the region
of relevance to charm production
at RHIC and LHC energies.  In addition, the onset of
saturation in the ratio of structure functions for
Xe to Deuterium is observed at much smaller value of x
than predicted by the parton recombination model.$^{10}$
Thus, it is not surprising that
even the best fit of Eq. (4)
overestimates the observed shadowing
effect by about
$15\%$.
Consequently
charm production
calculated with this shadowing function
would be underestimated
by about $40\%$.

We calculate charm production using
the
shadowing function that has recently been
proposed as the best fit to
EMC, NMC and E665 data$^{10}$ and is given by$^{11}$
\begin{equation}
R(x,A)
=\left\{\begin{array}{ll}
\alpha_3 -\alpha_4 x & x_0 <x\leq 0.6 \\
(\alpha_3 -\alpha_4 x_0)\frac{1+k_q \alpha_2 ({1/x}-1/x_{0})}
{1+k_q A^{\alpha_1}({1/x}-1/x_{0})}}
&x\leq x_{0}\\
\end{array}
\right
\end{equation}
The values for the parameters $k_q$, $\alpha_{1}$,
$\alpha_2$, $\alpha_3$ and $x_0$ can be found in Ref. 11.
\par
Presently
there is no theory which
can quantitatively describe the observed
nuclear shadowing effect.
Recent calculations of the
perturbative gluon shadowing indicate that non-perturbative
effects are large and can not be neglected.$^{12}$
\par
By integrating Eq. (2) and multipling by $T_{AA}(0)$ (for central
collisions) or
by $A^2/\sigma_{in}^{AA}$ (for inelastic collisions),
we obtain rapidity distribution,
transverse momentum distribution and the total cross section for the
inclusive charm quark
production in AA collisions.  Comparison of
our results for the total
charm cross section with the low-energy hadronic and nuclear
data ($20GeV\leq \sqrt s \leq 55GeV$)$^{13}$
indicate
very good agreement for the renormalization scale $Q=m_c$.$^{14}$
When $Q=2m_c$ is used, our cross sections
are slightly smaller than the measured
values.

\section{Results}

In our calculation we
use the two-loop evolved parton structure functions, MRS-S0
with
$\Lambda_5=140MeV$.$^{15}$
By using two other sets of structure functions, MRS-D0 and
MRS-D--,$^{15}$
we find that theoretical uncertainty due to the choice of
the nucleon structure function is only $10\%$ at RHIC.
This is not surprising because
the average $x$-value probed with charm production at $\sqrt s=200GeV$
is about $10^{-2}$, which is still within the range of $x$ for which
there is deep inelastic lepton-nucleon scattering data.
However, at LHC energies, where charm production is probing very
small values of $x$ ($x_{av} \sim 10^{-4}$),
the results are very sensitive to the gluon
structure function in the low-x region, far below the range
that has currently been explored by
deep-inelastic scattering experiments.$^{10}$  In particular, we
find that the total cross section for charm production is
about a factor of six larger
when we use MRS-D-- structure function, which
has ``singular'' behavior at low x (i.e. $G(x)\sim x^{-1.5}$),
instead of the MRS-S0 set.$^{15}$  In case of the rapidity
distribution, the results for charm production
obtained using MRS-D- structure
function are
about factor of $4.6$ larger for $\vert y \leq 1$.
In addition,
at the LHC the
theoretical uncertainty due to the choice of scale is about
$50\%$.
\par
Our results for the rapidity distribution of
inclusive charm production in central Au-Au collisions at RHIC and
LHC are
presented in Fig. 1 (solid
line).  We also show the rapidity distribution when
nuclear shadowing is not included (dotted line) and
 the
leading-order results with shadowing (short-dashed line) and without
shadowing (long-dashed line).
We note the the {\it shape} of the rapidity distribution does not
seem to be affected by
the next-to-leading corrections or
by the nuclear shadowing effect.
We find that, in the central rapidity region,
the number of charm
quark pairs produced per unit rapidity in central Au-Au collisions
at RHIC and LHC
is
$0.9$ and $4$ respectively.

\vspace*{7.2cm}
\noindent
{\small Fig. 1.
Rapidity distribution
of inclusive charm quark
production
at RHIC (a) and LHC (b).}
\vskip 0.2true in
\par
In hadronic collisions, one usually defines the ``K-factor''
as a measure of the size of higher-order corrections.  Here we
define the {\it effective} K-factor for {\it nuclear} collisions
as a ratio of the particular distribution
to the leading-order distribution without
any nuclear effects.
We find that in
hadronic collisions
K-factor is about $2$ (at RHIC) and $3$ (at LHC).
The nuclear shadowing effect suppresses production of charm quarks
by
about $30\%$ (at RHIC) and $53\%$ (at LHC), giving
the effective,
{\em nuclear}  K-factor of
about $1.4$ at both energies,
in the central rapidity region ($\vert y\vert \leq 3$).

In Fig. 2 we present our results for the transverse momentum
distribution of the
charm quark produced in Au-Au collisions at RHIC and LHC (solid line).
We find
that both higher-order correlations and the nuclear effects change
the shape of
$p_{T}$
distribution.  At RHIC,
the nuclear shadowing effect is much stronger at
low $p_T$ (about $40\%$ effect), while at $p_T=6GeV$ it reduces the
cross section by only
few percent.
The next-to-leading order corrections give a factor of $1.7$
increase at low $p_T$ and about factor of $3$ at $p_T=6GeV$.  These
two effects together result in
effective K-factor increasing from $1$ at $p_T=1GeV$ to
about $3$
at $p_T=6GeV$.  At large $p_T$, where nuclear shadowing effects are
negligible, we expect K-factor to approach its hadronic value.
At LHC, the nuclear shadowing effect is about
$60\%$ at low $p_T$ and $40\%$ at $p_T=6GeV$.  The nuclear K-factor
increases from $0.4$ at $p_T=1GeV$ to about $7$ at $p_T=6GeV$.
We find similar behavior
of the K-factor for $x_F$ distribution, namely
its strong dependence on $x_F$.$^{14}$

By integrating differential distributions over the phase space
we obtain the total number of charm quark pairs produced.
For central Au-Au collisions at RHIC (LHC)
we get about $4$ ($40$) charm quark pairs produced per event.

\vspace*{7.2cm}
\noindent {\small Fig. 2
Transverse momentum distribution
of inclusive charm quark
production
at RHIC (a)
\par
and LHC (b).}

\section{Conclusion}

\par
To conclude, we
have presented the complete next-to-leading order
calculation of the
differential and total inclusive cross sections
for the charm production in hadronic and nuclear
collisions at RHIC and LHC energies.
We have shown that
both higher-order contributions and the
nuclear shadowing effect
are large and can not be
neglected. In the central region
of the rapidity distribution,
the higher-order contributions increase the cross section
by a factor of $2$ ($3$), while the nuclear shadowing effect result in
additional decrease
of
about $1.4$ ($2.1$).
These two effects together result in
{\it nuclear} K-factor of
about $1.4$ at both energies in the central
rapidity region.  On the other hand,
in case of
the
$p_T$
distribution, the  K-factor changes from $1$ at $p_T=1GeV$ to
$3$ at $p_T=6GeV$ at RHIC and from
$0.4$ at $p_T=1GeV$ to $7$ at $p_T=6GeV$ at LHC.
We have found that at RHIC and LHC energies
the dominant subprocess for charm
production is gg fusion ($\geq 95\%$).
Therefore, future measurements of charm production in p-p and
A-A collisions at RHIC and LHC energies,
in addition to being a test of perturbative QCD,
could provide valuable information about presently unknown
$A$-dependence and $x$-dependence
of the gluon density in a nucleus.

Finally, we make a remark on the possibility of detecting  a
signal for
the formation of quark-gluon plasma via enchanced charm production
in heavy-ion collisions at RHIC.  We have found that
$0.9$
open charm quark pairs per unit rapidity (in the central region)
will be produced
in central Au-Au collisions via hard parton-parton scatterings.
This result seems to indicate that
the initial temperature of the quark-gluon plasma formed in Au-Au
collisions at RHIC would need to be unrealistically high$^5$ to
overcome the QCD background.  However, further theoretical
and experimental work is
needed in better understanding of
the nuclear shadowing effects on
the gluon density,
before definite conclusion can be made.

\section{Acknowledgements}

The work presented here was done in collaboration with P. Valerio.
We are grateful to M. Mangano
for providing us with
the fortran routines for calculating double differential
distributions in hadronic collisions and for many
helpful comments.
This work was supported in part through
U.S. Department of Energy Grants Nos.
DE-FG03-93ER40792 and
DE-FG02-85ER40213.
\vskip 0.2 true in
\bibliographystyle{unsrt}

\begin{thebibliography}{99.}
\bibitem{1} I. Sarcevic and P. Valerio, University of Arizona preprint
AZPH-TH/93-13, submitted to {\em Phys. Lett.} {\bf B}.
\bibitem{2}
For a recent review see, {\em Quark Matter '93}
ed. H. A. Gustaffson,
{\em Nucl. Phys.} {\bf A566} (1994).
\bibitem{3}
For example, see
B.  Muller, in {\em Particle Production in Highly Excited Matter}, eds.
H. Gutbrod and J. Rafelski (Plenum, New York, 1993).
\bibitem{4}  For a recent review see,
S. Gavin, these {\em Proceedings}.
\bibitem{5} E. Shuryak, {\em Phys. Rev. Lett.} {\bf
68}
(1992) 3270; K. Geiger,
{\em Phys. Rev.} {\bf D48} (1993) 4129.
\bibitem{6}  M. L. Mangano, P. Nason and G. Ridolfi, {\em
Nucl. Phys.}
{\bf B373} (1992) 295.
\bibitem{7} K. J. Eskola, K. Kajantie and J. Lindfors, {\em
Nucl. Phys.} {\bf B323}
(1989) 37.
\bibitem{8} A. Bohr and B. R. Mottelson,
{\em Nuclear Structure I} (Benjamin, New
York, 1969) pp. 160, 223.
\bibitem{9} J. Qiu, {\em Nucl. Phys.}
{\bf B291} (1987) 746; K. J. Eskola,
{\em Nucl. Phys.} {\bf B400} (1993)  240.
\bibitem{10}  NMC Collaboration, P. Amaudruz {\it et al.},
{\em Z. Phys.} {\bf C51} (1991) 387;
E665 Collaboration, M. R. Adams {\it et al.},
{\em Phys. Rev. Lett.} {\bf {68}} (1992) 3266; EMC Collaboration,
J. Ashman
{\it et al.}, {\em Phys. Lett.} {\bf B202} (1988) 603;
M. Arnedo {et al.}, {\em Phys. Lett.} {\bf B211} (1988) 493.
\bibitem{11}  C. J. Benesh, J. Qiu and J. P. Vary,
Los Alamos preprint, LA-UR-94-784.
\bibitem{12}  K. J. Eskola, J. Qiu and X.-N. Wang,
{\em Phys. Rev. Lett.} {\bf 72}, 36 (1994).
\bibitem{13} S. Aoki {\it et al.}, {\em Phys. Lett.}
{\bf B224} (1989) 441;
S. P. K. Tavernier, {\em Rep. Prog. Phys.} {\bf 50} (1987) 1439;
S. Barlag {\it et al.}, {\em Z. Phys.} {\bf C39} (1988) 451.
\bibitem{14}  I. Sarcevic and P. Valerio, University of
Arizona preprint, AZPH-TH/94-20.
\bibitem{15}
A. D. Martin, W. J. Sterling and R. G. Roberts,
{\em Phys. Rev.}
{\bf D47} (1993) 867.
\end{thebibliography}

\vskip 0.15true in
\noindent
\vfil\eject
\end{document}